# Effect of Bohm potential on a charged gas

**D. Mostacci**[*], **V. Molinari, F. Pizzio**

*Laboratorio di Ingegneria Nucleare di Montecuccolino*
*Via dei Colli 16, I-40136 Bologna (ITALY)*
*Alma Mater Studiorum - Università di Bologna (Bologna, Italy)*

**ABSTRACT**

Bohm's interpretation of Quantum Mechanics leads to the derivation of a Quantum Kinetic Equation (QKE): in the present work, propagation of waves in charged quantum gases is investigated starting from this QKE. Dispersion relations are derived for fully and weakly degenerate fermions and bosons (these latter above critical temperature), and the differences underlined. Use of a kinetic equation permits investigation of "Landau-type" damping: it is found that the presence of damping in fermion gases is dependent upon the degree of degeneracy, whereas it is always present in boson gases. In fully degenerate fermions a phenomenon appears that is akin to the "zero sound" propagation.

**I. Introduction**

In recent years much work has focused on transport processes that need to be treated from a quantum mechanical point of view. Just to name a few, electron gas in metals, miniaturisation of electronic devices, dense astrophysical systems, dense plasmas, laser-matter interaction [1, 2, 3, 4, 5, 6]. Much of the recent work, however, has been conducted in the framework of the fluid model, i.e., through macroscopic equations. The present work does not analyse a specific problem, but rather its aim is to investigate quantum effects on certain fundamental phenomena, like wave propagation and Landau damping in fermion and in boson systems: to this end, a kinetic equation approach is needed.

---

[*] Corresponding author - fax: +39-051-6441747; tel.: +39-051-6441711; e-mail: *domiziano.mostacci@unibo.it*

Kinetic equations incorporating quantum effects can be derived from a Bohm potential point of view [7, 8]. Bohm potential is given by the following expression for a system of N identical particles of mass m:

$$U(\mathbf{r}_1,...,\mathbf{r}_N,t) = -\frac{\hbar^2}{2mR}\sum_{l=1}^{N}\nabla_l^2 R_N(\mathbf{r}_1,...,\mathbf{r}_N,t) \tag{1}$$

where $R_N(\mathbf{r}_1,...,\mathbf{r}_N,t)$, a real function, is the modulus of the N-particle wave function $\Psi_N$:

$$\Psi_N(\mathbf{r}_1,...,\mathbf{r}_N,t) = R_N(\mathbf{r}_1,...,\mathbf{r}_N,t)\exp\left\{\frac{iS_N(\mathbf{r}_1,...,\mathbf{r}_N,t)}{\hbar}\right\} \tag{2}$$

Disregarding the effects of correlation in calculating the contribution of quantum potential as, e.g., in the Hartree method [9],

$$R_N(\mathbf{r}_1,\mathbf{r}_2\cdots\mathbf{r}_N,t) = R(\mathbf{r}_1,t)R(\mathbf{r}_2,t)\cdots R(\mathbf{r}_N,t) \tag{3}$$

where $R(\mathbf{r},t)$ is the modulus of the single particle wave function $\Psi$ and the subscripts 1,…,N refer to particle 1, …, N respectively. A quantum kinetic equation (QKE) can be derived [8] in the "traditional" form:

$$\frac{\partial f}{\partial t} + \mathbf{v}\cdot\frac{\partial f}{\partial \mathbf{r}} + \frac{\mathbf{F}_C + \mathbf{F}_Q}{m}\cdot\frac{\partial f}{\partial \mathbf{v}} = \left(\frac{\partial f}{\partial t}\right)_{coll} \tag{4}$$

where f is the single distribution function, $\left(\frac{\partial f}{\partial t}\right)_{coll}$ is the applicable (as yet unspecified) collision term, the quantum force $\mathbf{F}_Q$ derives from Bohm potential:

$$\mathbf{F}_Q = -\frac{\partial}{\partial \mathbf{r}_l}U_B = \frac{\partial}{\partial \mathbf{r}_l}\left(\frac{\hbar^2}{2m^2}\frac{\nabla^2 R}{R}\right) \tag{5}$$

and $\mathbf{F}_C$ is the classical force term.

If in particular charged particles are considered, in the absence of external fields, the mutual electrostatic force will need to be considered: this is effected through a self-consistent field $\mathbf{E}$, and if near collisions can be neglected the Quantum Vlasov Equation (QVE) is obtained as

$$\frac{\partial f}{\partial t} + \mathbf{v}\cdot\frac{\partial f}{\partial \mathbf{r}} + \frac{q\mathbf{E} + \mathbf{F}_Q}{m}\cdot\frac{\partial f}{\partial \mathbf{v}} = 0 \tag{6}$$

where a second equation will be needed to determine the electric field. A usual choice is Gauss equation:

$$\frac{\partial}{\partial \mathbf{r}_1} \cdot \mathbf{E} = \frac{q}{\varepsilon_0} n \qquad (7)$$

where q is the charge on the particles and $\varepsilon_0$ is the permittivity of free space.

Aim of the present work is to apply the above description to derive dispersion relations for waves of small amplitude in systems of boson and of fermion gases. This will be presented in the following sections.

**II. Wave propagation**

Consider a small perturbation in a gas of (possibly) electrically charged particles, with no external forces, in some equilibrium described by a homogeneous, uniform distribution function $f_0(\mathbf{v})$, and in which $\mathbf{E} = 0$. Let the small perturbation be described as follows:

$$f(z, \mathbf{v}, t) = f_0(\mathbf{v}) + \varphi(z, \mathbf{v}, t) \qquad (8)$$

and then the particle density n writes as

$$n(z,t) = \int_{\mathfrak{R}^3} f(z,\mathbf{v},t) d_3 v = \int_{\mathfrak{R}^3} f_0(\mathbf{v}) d_3 v + \int_{\mathfrak{R}^3} \varphi(z,\mathbf{v},t) d_3 v = n_0 + \nu(z,t) \qquad (9)$$

In the present case of variation only along z

$$\mathbf{E}(z,t) = E(z,t)\hat{\mathbf{z}} \qquad (10)$$

$$\mathbf{F}_Q = F_Q \hat{\mathbf{z}} = \frac{d}{dz}\left(\frac{\hbar^2}{2m} \frac{d^2 R}{dz^2} \frac{1}{R}\right)\hat{\mathbf{z}} \qquad (11)$$

It is clear that possible wave propagation under these condition will be, if any, along the z direction. The starting point is (5), particularised for the present case:

$$\frac{\partial f}{\partial t} + v_z \frac{\partial f}{\partial z} + \frac{qE + F_Q}{m} \frac{\partial f}{\partial v_z} = 0 \qquad (12)$$

Keeping in mind the above definition of $F_Q$, and the fact that the density n is proportional to $R^2$, the square of the modulus of the single particle wave function, introducing the perturbation and neglecting all terms of order higher than first the quantum force becomes

$$F_Q = \frac{\hbar^2}{4mn_0} \frac{\partial^3 \nu}{\partial z^3} \qquad (13)$$

Effect of Bohm potential on a charged gasLikewise, introducing the perturbation in the QVE and using the above expression for the perturbed quantum force, neglecting again all terms of order higher than first, the linearised QVE is obtained, together with its companion Gauss equation,

$$\frac{\partial \varphi}{\partial t} + v_z \frac{\partial \varphi}{\partial z} + \frac{qE}{m}\frac{\partial f_0}{\partial v_z} + \frac{\hbar^2}{4m^2 n_0}\frac{\partial f_0}{\partial v_z}\frac{\partial^3 \nu}{\partial z^3} = 0 \tag{14}$$

$$\frac{\partial}{\partial z}E = \frac{q}{\varepsilon_0}\nu = \frac{q}{\varepsilon_0}\int_{\Re^3}\varphi\, d_3 v \tag{15}$$

where $f_0$ is the Fermi-Dirac distribution function for fermions and that of Bose-Einstein for bosons.

Taking a Fourier transform with respect to the space variable and a Laplace transform with respect to time of eqs. (14) and (15) and rearranging, calling $\Psi(k,\mathbf{v},s)$ the double transform of $\varphi(z,\mathbf{v},t)$, $\mathcal{N}(k,s)$ that of the density $\nu(z,t)$ and $\mathcal{E}(k,s)$ that of the electric field $E(z,t)$

$$\Psi(k,\mathbf{v},s) + \left[\frac{q}{m}\mathcal{E}(k,s) - ik^3 \frac{\hbar^2}{4m^2 n_0}\mathcal{N}(k,s)\right]\frac{1}{s+ikv_z}\frac{\partial f_0(\mathbf{v})}{\partial v_z} = \frac{\tilde{\varphi}(k,\mathbf{v},t=0)}{s+ikv_z} \tag{16}$$

$$ik\mathcal{E}(k,s) = \frac{q}{\varepsilon_0}\mathcal{N}(k,s) \tag{17}$$

where $\tilde{\varphi}(k,\mathbf{v},t=0)$ is the Fourier transform of the initial perturbation $\varphi(z,\mathbf{v},t=0)$. Now, introducing (17) into (16) and recalling that

$$\mathcal{N}(k,s) = \int_{\Re^3}\Psi(k,\mathbf{v},s)d\mathbf{v} \tag{18}$$

the result can be integrated throughout over velocity space yielding

$$\mathcal{N}(k,s)\left\{1 - i\left[\frac{1}{k}\frac{q^2}{m\varepsilon_0} + k^3\frac{\hbar^2}{4m^2 n_0}\right]\int_{\Re^3}\frac{\partial f_0(\mathbf{v})}{\partial v_z}\frac{d\mathbf{v}}{s+ikv_z}\right\} = \int_{\Re^3}\tilde{\varphi}(k,\mathbf{v},t=0)\frac{d\mathbf{v}}{s+ikv_z} \tag{19}$$

or, after rearranging and performing the integration over $v_x$ and $v_y$,

$$\mathcal{N}(k,s) = \frac{\displaystyle\int_{-\infty}^{+\infty}\tilde{\varphi}_z(k,v_z)\frac{dv_z}{s+ikv_z}}{1 - i\left[\dfrac{1}{k}\dfrac{q^2}{m\varepsilon_0} + k^3\dfrac{\hbar^2}{4m^2 n_0}\right]\displaystyle\int_{-\infty}^{+\infty}\frac{\partial f_z(v_z)}{\partial v_z}\frac{dv_z}{s+ikv_z}} \tag{20}$$

where the following notations have been adopted:

$$\tilde{\varphi}_z(k, v_z) = \iint_{\Re^2} \tilde{\varphi}(k, \mathbf{v}, t=0) dv_x dv_y \tag{21}$$

$$f_z(v_z) = \iint_{\Re^2} f_0(\mathbf{v}) dv_x dv_y \tag{22}$$

The integrals in (20) are to be calculated in the complex w plane, where w is a complex variable of which $v_z$ is the real part. (20) is then written more accurately as

$$\mathcal{N}(k,s) = \frac{\frac{1}{ik}\int_\Gamma \tilde{\varphi}_z(k,w)\frac{dw}{w-i\frac{s}{k}}}{1-\frac{1}{k^2}\left(\Omega_p^2 + k^4\frac{\hbar^2}{4m^2}\right)\frac{1}{n_0}\int_\Gamma \frac{df_z(w)}{dw}\frac{dw}{w-i\frac{s}{k}}} \tag{23}$$

where the path $\Gamma$ is the straight line that lies on the real axis, and the following symbol has been used:

$$\Omega_p^2 = \frac{q^2 n_0}{m\varepsilon_0} \tag{24}$$

In a plasma, the above expression is the plasma frequency. Laplace antitransformation of (16) is rather involved, however it is not indispensable if interest is focused on the asymptotic solution for large t [10, 11, 12, 13, 7], for then the evolution is dominated by the rightmost zero of the denominator in (23), and if its imaginary part is negative this will give rise to a "Landau-type" damping. The problem is hence reduced to finding the solutions of the following equation

$$1 - \frac{1}{k^2}\left(\Omega_p^2 + k^4 \frac{\hbar^2}{4m^2}\right)\frac{1}{n_0}\int_\Gamma \frac{df_z(w)}{dw}\frac{dw}{w-i\frac{s}{k}} = 0 \tag{25}$$

The derivative inside the integral is different for fermions and bosons

$$\frac{df_z(w)}{dw} = -2\pi w \gamma \frac{m^3}{h^3}\frac{1}{\frac{1}{\alpha}\exp\left\{\frac{mw^2}{2K_B T}\right\} \pm 1} \tag{26}$$

where the upper sign holds for fermions, and the lower for bosons; here $\gamma$ is the statistical weight, $K_B$ Boltzmann constant, and the fugacity $\alpha$ is connected to the chemical potential $\mu$ by the usual relation $\alpha = \exp[\mu/K_B T]$.

## III. Dispersion relations

### III.1 Fully degenerate fermions

In the limiting case of full degeneracy, the Fermi-Dirac distribution function approaches the value

$$f_{FD}(\mathbf{v}) = \frac{\gamma m^3}{h^3} U(v_F^2 - v^2) \tag{27}$$

$v_F$ being the velocity corresponding to the Fermi energy:

$$v_F = \left(\frac{3n_0 h^3}{4\pi\gamma m^3}\right)^{\frac{1}{3}} \tag{28}$$

The derivative of the reduced distribution of (22) becomes for this case

$$\frac{df_z(v_z)}{dv_z} = -2\pi \frac{\gamma m^3}{h^3} v_z U(v_F^2 - v_z^2) \tag{29}$$

Then the integral in (25) can be calculated, starting with the Cauchy principal part:

$$\int \frac{df_z(v_z)}{dv_z} \frac{dv_z}{v_z - i\frac{s}{k}} = 4\pi \frac{\gamma m^3}{h^3}\left[\frac{s}{k}\operatorname{arctg}\frac{kv_F}{s} - v_F\right] \tag{30}$$

where the arctg function is defined here strictly in the $\left[-\frac{\pi}{2}, \frac{\pi}{2}\right]$ interval; and the residual

$$\operatorname{Re}_{i\frac{s}{k}}\left[\frac{df_z(v_z)}{dv_z}\frac{1}{v_z - i\frac{s}{k}}\right] = 4\pi^2 \frac{\gamma m^3}{h^3}\frac{\eta + i\omega}{k} U(k^2 v_F^2 + \eta^2 - \omega^2) \tag{31}$$

having set $s = \eta + i\omega$; (25) becomes

$$1 = \frac{1}{k^2}\left(\Omega_p^2 + k^4 \frac{\hbar^2}{4m^2}\right)\frac{3}{v_F^3}\left[\frac{s}{k}\operatorname{arctg}\frac{kv_F}{s} - v_F + \pi\frac{s}{k}U(k^2 v_F^2 + \eta^2 - \omega^2)\right] \tag{32}$$

Writing the real and imaginary parts of (32) the following system is obtained

$$1 = \frac{3}{v_F^3}\left(\Omega_p^2 + \frac{\hbar^2 k^4}{4m^2}\right)\frac{v_\Phi}{k^2}\left[\frac{\varepsilon}{2}\operatorname{arctg}\frac{2r\varepsilon}{1+\varepsilon^2 - r^2} - \frac{1}{4}\ln\frac{(1-r)^2 + \varepsilon^2}{(1+r)^2 + \varepsilon^2} - r + \pi\varepsilon\, U(r^2 + \varepsilon^2 - 1)\right] \tag{33}$$

$$0 = \left[\frac{1}{2}\operatorname{arctg}\frac{2r\varepsilon}{1+\varepsilon^2 - r^2} + \frac{\varepsilon}{4}\ln\frac{(1-r)^2 + \varepsilon^2}{(1+r)^2 + \varepsilon^2} + \pi U(r^2 + \varepsilon^2 - 1)\right] \tag{34}$$

where $v_\Phi$ has been written for the ratio $\omega/k$, i.e., the phase velocity, and r and $\varepsilon$ have been written as a shorthand:

$$r = \frac{v_F}{v_\Phi}; \qquad \varepsilon = \frac{\eta}{\omega} \qquad (35)$$

Notice preliminarily that $\varepsilon = 0$ solves identically (34) whenever the Heavyside function vanishes, i.e. whenever $r^2 + \varepsilon^2 < 1$. Consider first the case $r \ll 1$, i.e., $v_\Phi \gg v_F$, and assume that also $\varepsilon \ll 1$ (for large values of $\varepsilon$ the wave is overdamped and there is no propagation): then the Heavyside function in both the above equations vanishes identically, and making $\varepsilon = 0$ in (33) and keeping only the terms of lower order in r:

$$\frac{3}{v_F^3}\left(\Omega_p^2 + \frac{\hbar^2 k^4}{4m^2}\right)\frac{v_\Phi}{k^2}\left[r + \frac{1}{3}r^3 - r\right] = 1 \qquad (36)$$

yielding

$$\omega^2 = \Omega_p^2 + \frac{\hbar^2 k^4}{4m^2} \qquad (37)$$

The same expression was obtained, e.g., in [3] (for vanishing streaming velocity) from a Wigner equation treatment.

If a further term is retained in the expansion of the logarithm, i.e., the term in $r^5$, the dispersion relation becomes

$$\omega^2 = C_1 \frac{1 + \sqrt{1 + \frac{12}{5}\frac{k^2 v_F^2}{C_1}}}{2} \qquad (38)$$

having set

$$C_1 = \Omega_p^2 + \frac{\hbar^2 k^4}{4m^2} \qquad (39)$$

For small enough values of k, such that $k^2 v_F^2 \ll C_1$ the above equation yields

$$\omega^2 = \Omega_p^2 + \frac{3}{5}k^2 v_F^2 + \frac{\hbar^2}{4m^2}k^4 \qquad (40)$$

The above expression, neglecting the third term stemming from Bohm potential reverts to the expression reported by Landau [14, 13]. Also of interest the experimental trend observed by [15], exhibiting the same behaviour as in (40), se also [6].

Consider now the case $v_\Phi > v_F$ only slightly, assuming once more $\varepsilon \ll 1$: again, the heavyside function vanishes, leaving $\varepsilon = 0$. Then (33) becomes

$$1 = \frac{3}{v_F^3}\left(\Omega_p^2 + \frac{\hbar^2 k^4}{4m^2}\right)\frac{v_\Phi}{k^2}\left[-\frac{1}{4}\ln\frac{(1-r)^2}{(1+r)^2}-r\right] \quad (41)$$

and can be solved readily to yield

$$\omega = kv_F\left[1 + 2\exp\left\{-2\frac{k^2 v_F^2}{3}\left(\Omega_p^2 + \frac{\hbar^2 k^4}{4m^2}\right)^{-1} - 2\right\}\right] \quad (42)$$

Here $v_\Phi > v_F$, consistently with the initial assumption. It may be noticed that for the case of neutral fermions, the above expression parallels the zero sound propagation [16, 17, 13], approaching $v_\Phi = v_F$ for $k \to 0$.

A prominent feature of the above results is the absence of damping, $\varepsilon = 0$: this was to be expected since, as discussed by Landau [14], when the phase velocity is larger than the Fermi velocity of a fully degenerate fermion system, there can be no resonant particles, i.e., particles with velocity equal to that of the propagating wave. Therefore no energy can be exchanged between the wave and the particles.

### III.2 Bosons above $T_c$ and weakly degenerate fermions

Both case correspond to values of the parameter $\alpha < 1$, therefore the ratio on the rhs of (26) can be expanded in series of exponentials, and then integration performed leading to the following dispersion relation, paralleling [7]:

$$\left(\frac{\Omega_p^2}{k^2} + \frac{\hbar^2}{4m^2}k^2\right)\sqrt{\frac{2\pi K_B T}{m}}\left[\sum_{j=1}^{\infty}(\mp 1)^{j-1}\frac{\alpha^j}{\sqrt{j}}\left[\sqrt{\pi}\sqrt{j}\theta e^{j\theta^2}\text{erfc}(\sqrt{j}\,\theta)-1\right] + \frac{2\sqrt{\pi}\vartheta}{\frac{1}{\alpha}e^{-\vartheta^2}\pm 1}\right] = \frac{2}{3}v_{ch}^3 \quad (43)$$

where the characteristic velocity $v_{ch}$ is given by the following expression

$$v_{ch} = \left(\frac{3n_0 h^3}{4\pi\gamma m^3}\right)^{\frac{1}{3}} \quad (44)$$

which is seen to be the Fermi velocity in the case of fermions, and

$$\vartheta = \sqrt{\frac{m}{2K_B T}}\frac{s}{k} \quad (45)$$

Following further [7], the case $|\vartheta| \gg 1$ can be analysed taking the following expansion for the complementary error function:

$$\sqrt{\pi}\, z\, e^{z^2}\text{erfc}(z) - 1 = -\frac{1}{2z^2} + \frac{3}{4z^4} + O[z^{-6}] \quad (46)$$

Yielding

$$\left(\frac{\Omega_p^2}{k^2} + \frac{\hbar^2}{4m^2}k^2\right)\left[\sqrt{\pi}\left(\frac{2K_BT}{m}\right)^{\frac{3}{2}}\zeta_{\frac{3}{2}}^{\pm}(\alpha)\left[\left(v_{th}^{\pm}\right)^2\frac{k^4}{s^4} - \frac{k^2}{s^2}\right] + \frac{4\pi\frac{s}{k}}{\frac{1}{\alpha}e^{-\frac{m}{2K_BT}\frac{s^2}{k^2}} \pm 1}\right] = \frac{4}{3}v_{ch}^3 \quad (47)$$

where the thermal velocity is given by [18, 19]

$$\left(v_{th}^{\pm}\right)^2 = \frac{3K_BT}{m}\frac{\zeta_{\frac{5}{2}}^{\pm}(\alpha)}{\zeta_{\frac{3}{2}}^{\pm}(\alpha)} \quad (48)$$

and the $\zeta$ functions are defined as [18, 19]:

$$\zeta_r^{\pm}(\alpha) = \sum_{j=1}^{\infty}(\mp 1)^{j-1}\frac{\alpha^j}{j^r} \quad (49)$$

Also [18],

$$\zeta_{\frac{3}{2}}^{\pm}(\alpha) = \frac{n_0}{\gamma}\frac{h^3}{(2\pi K_BT)^{\frac{3}{2}}} \quad (50)$$

Upon multiplying by $s^4$ and considering the real part, a biquadratic equation for $\omega$ is obtained as follows

$$\omega^4 - \omega^2\left(\Omega_p^2 + \frac{\hbar^2}{4m^2}k^4\right) - k^2\left(v_{th}^{\pm}\right)^2\left(\Omega_p^2 + \frac{\hbar^2}{4m^2}k^4\right) = 0 \quad (51)$$

yielding the solution

$$\omega^2 = \frac{1}{2}\left(\Omega_p^2 + \frac{\hbar^2}{4m^2}k^4\right) + \frac{1}{2}\sqrt{\left(\Omega_p^2 + \frac{\hbar^2}{4m^2}k^4\right)^2 + 4k^2\left(v_{th}^{\pm}\right)^2\left(\Omega_p^2 + \frac{\hbar^2}{4m^2}k^4\right)} \quad (52)$$

The above result was obtained upon the condition $|\vartheta| \gg 1$, hence $k^2\left(v_{th}^{\pm}\right)^2 \ll \Omega_p^2$; therefore a further simplification can be obtained approximating the square root:

$$\omega^2 = \left(\Omega_p^2 + \frac{\hbar^2}{4m^2}k^4\right) \times \left[1 + \frac{k^2\left(v_{th}^{\pm}\right)^2}{\left(\Omega_p^2 + \frac{\hbar^2}{4m^2}k^4\right)}\right] \quad (53)$$

or simply

$$\omega^2 = \Omega_p^2 + k^2 \left(v_{th}^{\pm}\right)^2 + \frac{\hbar^2}{4m^2} k^4 \tag{54}$$

This result can be seen to revert to the classical dispersion relation for Langmuir waves in plasmas in the limit $\hbar \to 0$.

Considering fermions, comparison of the above expression with (40) shows a marked similarity. The structure of the two equations is the same, the only difference being in the velocity in the second term: the thermal velocity for the weakly degenerate case and the Fermi velocity if degeneracy is full.

As far as bosons, it must be stressed that (54) was obtained taking in no account binary interactions among particles, but rather considering bosons subject only to Bohm potential and self-consistent electric field. The model would clearly be inadequate to describing a liquid, e.g. liquid helium, where interparticle interaction is strong enough to modify substantially the physical framework. However, in the gases above critical temperature considered here interactions are too weak to produce those feature typical of boson liquids, for instance, the rotonic minimum in the dispersion relation [20,21,22].

**IV. CONCLUSIONS**

If quantum effects are represented through Bohm potential, a QKE is derived along the lines of classical statistical mechanics. Waves can be studied with this QKE: in the present work, this procedure is applied to investigate wave propagation in fermion and in boson gaseous systems, possibly charged, either fully or weakly degenerate. As should be expected, the dispersion relations obtained are complex: the real and imaginary parts constitute a system of two equations in the two unknown ω and η, the frequency of the wave and the damping factor. In principle, then, these two quantities can be expressed as functions of the wave number k. In the case of fermions, there is a marked difference between weak and full degeneracy: in the former case the thermal velocity appears in the dispersion relation (see (54)), whereas in the latter the Fermi velocity takes its place ((40)), therefore there is propagation even at vanishing temperature (zero sound). Furthermore, it is found that no Landau damping is present in this latter case. This is to be contrasted with the behaviour of bosons for which a Landau damping is always present. The ratio of phase velocity to Fermi velocity plays an important role in the behaviour of fully degenerate fermions, as the dispersion relation is widely different.